\documentclass[aps,prl,twocolumn,preprintnumbers,superscriptaddress,
              showpacs,nofootinbib]{revtex4}
\newcommand{\PRE}[1]{}       

\usepackage{bm} \usepackage{epsfig}

\newcommand{\mweak}{M_{\text{weak}}}
\newcommand{\vmax}{v_{\text{max}}}

\newcommand{\OmegaWIMP}{\Omega_{\text{WIMP}}}
\newcommand{\OmegaSWIMP}{\Omega_{\text{SWIMP}}}
\newcommand{\mplanck}{M_{\text{Pl}}} 
 
\newcommand{\sigmaA}{\sigma_{\text{A}}} 
\newcommand{\sigmaan}{\sigma_{\text{an}}} 
\newcommand{\sigmatot}{\sigma_{\text{tot}}} 

\newcommand{\OmegaDM}{\Omega_{\text{DM}}}
\newcommand{\ifb}{\text{fb}^{-1}} 
\newcommand{\iab}{\text{ab}^{-1}}

\newcommand{\gev}{\text{GeV}} 
\newcommand{\tev}{\text{TeV}}

\newcommand{\etal}{{\em et al.}}  
\newcommand{\eg}{{\em e.g.}}

\newcommand{\eqref}[1]{Eq.~(\ref{#1})}
\newcommand{\eqsref}[2]{Eqs.~(\ref{#1}) and (\ref{#2})}

\newcommand{\figref}[1]{Fig.~\ref{fig:#1}}

\newcommand{\WIMP}{\text{WIMP}} 
\newcommand{\SWIMP}{\text{SWIMP}}

\newcommand{\mL}{m_L} 
\newcommand{\mWIMP}{m_{\WIMP}} 
\newcommand{\mSWIMP}{m_{\SWIMP}}

\begin{document}

\preprint{UCI-TR-2005-11}

\title{
\PRE{\vspace*{1.5in}}
Lower Limit on Dark Matter Production at the Large Hadron Collider
\PRE{\vspace*{0.3in}}
}

\author{Jonathan L.~Feng}
\affiliation{Department of Physics and Astronomy, University of
California, Irvine, CA 92697, USA
\PRE{\vspace*{.5in}}
}

\author{Shufang Su}
\affiliation{Department of Physics, University of Arizona,
Tucson, AZ 85721, USA
\PRE{\vspace*{.5in}}
}

\author{Fumihiro Takayama%
\PRE{\vspace*{.2in}}
} 
\affiliation{Institute for High-Energy Phenomenology, Cornell
  University, Ithaca, NY 14853, USA
\PRE{\vspace*{.5in}}
}

\date{December 2005}

\begin{abstract}
\PRE{\vspace*{.3in}} We evaluate the prospects for finding evidence of
dark matter production at the Large Hadron Collider.  We consider
WIMPs and superWIMPs, weakly- and superweakly-interacting massive
particles, and characterize their properties through model-independent
parameterizations.  The observed relic density then implies lower
bounds on dark matter production rates as functions of a few
parameters.  For WIMPs, the resulting signal is indistinguishable from
background.  For superWIMPs, however, this analysis implies
significant production of metastable charged particles.  For natural
parameters, these rates may far exceed Drell-Yan cross sections and
yield spectacular signals.
\end{abstract}

\pacs{95.35.+d, 13.85.-t, 04.65.+e, 12.60.Jv}

\maketitle

The energy density of non-baryonic dark matter in the Universe is now
known to be~\cite{Spergel:2003cb}
\begin{equation}
\OmegaDM h^2 = 0.112 \pm 0.009 \ ,
\label{omega}
\end{equation}
where $\OmegaDM$ is this energy density in units of the critical
density, and $h \simeq 0.71$ is the normalized Hubble parameter. With
accompanying constraints, this implies that non-baryonic dark matter
makes up about a quarter of the energy density of the Universe.  Its
microscopic identity is at present unknown, however, and is one of the
outstanding questions in basic science today.

An exciting possibility is that particles that make up some or all of
dark matter may be produced at high-energy colliders, such as the
Tevatron at Fermilab or, beginning in 2007, the Large Hadron Collider
(LHC) at CERN.  Such prospects are particularly promising if dark
matter is composed of WIMPs or superWIMPs, weakly- or
superweakly-interacting massive particles, since these scenarios
require new particles with masses near $\mweak \sim 100~\gev$, the
scale to be probed in detail at the LHC.  Of course, the discovery of
dark matter signals requires not only that dark matter be produced at
colliders, but also that it be produced with rates and signatures that
allow it to be distinguished from background.  Remarkably, however,
WIMP and superWIMP production rates are also constrained by cosmology,
because dark matter densities are determined (in part) by thermal
freeze out in these scenarios. Since WIMPs and superWIMPs cannot have
relic densities in excess of $\OmegaDM h^2$, they must have
annihilated efficiently in the early Universe and consequently must be
produced efficiently at colliders.

In this paper, we analyze this argument quantitatively.  We begin by
characterizing the properties of WIMPs and superWIMPs through
model-independent parameterizations.  Our analysis is sufficiently
general to accommodate concrete realizations of WIMP or superWIMP dark
matter in supersymmetric, extra-dimensional, and many other
frameworks.  Using the observed relic density, we then determine, as
functions of the few parameters entering the analysis, the minimal
dark matter production cross section and its observable consequences.
This approach was applied in Ref.~\cite{Birkedal:2004xn} to obtain
conservative estimates for WIMP cross sections at the International
Linear Collider (ILC).  The results implied challenging rates for
discovery.  In this study, we focus on the more pressing case of the
LHC and also analyze superWIMP scenarios, for which our conclusions
will be far more promising.

In both the WIMP and superWIMP dark matter scenarios, the process of
freeze out plays a large role in determining the dark matter relic
density.  In WIMP scenarios, the dark matter is a stable, neutral WIMP
with mass $\mWIMP \sim \mweak$ and pair annihilation cross section
$\sigmaA \sim \alpha_{\text{weak}}^2 \mweak^{-2}$.  Beginning in
thermal equilibrium in the early Universe, WIMPs annihilate until they
become too dilute to find each other and ``freeze out'' at temperature
$T_F$.  Given the expansion rate determined by $\mplanck \simeq 1.2
\times 10^{19}~\gev$, along with $\mWIMP$ and $\sigmaA$ given above,
the thermal relic density $\OmegaWIMP h^2$ is automatically near the
observed $\OmegaDM h^2$ of \eqref{omega}.  In contrast to other dark
matter scenarios, there is no need to introduce new energy scales to
obtain the desired relic density.

SuperWIMP scenarios also begin with a weakly-interacting particle,
which we denote $L$, with mass $\mL \sim \mweak$ and $\sigmaA \sim
\alpha_{\text{weak}}^2 \mweak^{-2}$ that freezes out with density
$\Omega_L h^2 \sim \OmegaDM h^2$.  In contrast to WIMP scenarios,
however, $L$ particles are not stable, but metastable, and they
ultimately decay to superWIMPs, which form the dark matter we observe
today~\cite{Feng:2003xh}.  SuperWIMPs are neutral and stable, but
their interactions are much weaker than weak, typically gravitational.
The resulting dark matter density is, then,
\begin{equation}
\OmegaSWIMP h^2 = \frac{\mSWIMP}{\mL} \Omega_L h^2 \ .
\label{omegaswimp}
\end{equation}
If, as is often natural, $\mSWIMP \sim \mL$, superWIMPs are also
produced with relic densities of the right order of magnitude.  {\em A
priori}, $L$ particles may be either electrically charged or neutral.
However, in the well-motivated cases of gravitino and Kaluza-Klein
graviton superWIMPs~\cite{Feng:2003xh,Feng:2004mt,Fujii:2003nr}, the
neutral case is typically excluded by contraints from Big Bang
nucleosynthesis~\cite{Feng:2003xh,Feng:2004mt,Jedamzik:2004er}, and
the most motivated $L$ particles are charged sleptons and KK leptons.
In this study, we therefore consider only $L$ particles with charge
$\pm 1$.

To determine the production rates at colliders, we must first find
what annihilation cross sections are implied by the observed relic
density $\OmegaDM h^2$.  Let $X$ denote a generic particle that
freezes out, either the WIMP in WIMP scenarios, or $L$ in superWIMP
scenarios.  The total $X \bar{X}$ annihilation cross section is
\begin{equation}
\sigmatot = \textstyle{\sum_{ij}} \sigma(X\bar{X} \to ij; \hat{s}) \ ,
\end{equation}
where $\sqrt{\hat{s}}$ is the center-of-mass energy, and $i, j$ are
partons.  We parameterize the thermal average of $\sigmatot$ as
\begin{equation}
\langle \sigmatot v_X \rangle 
\equiv \sigmaan v_X^{2n} + {\cal O}(v_X^{2n+2}) 
\equiv \sigma_0 x^{-n} + {\cal O}(x^{-n-1}) ,
\label{sigmatot}
\end{equation}
where $v_X$ is the relative velocity of the initial $X$ and $\bar{X}$
particles in their center-of-mass frame, and $x \equiv m_X/T =
6/v_X^2$.  This expansion is valid in the common case where
annihilation is dominated by a single angular momentum component
($S$-wave for $n=0$, $P$-wave for $n=1$, etc.).  It is necessarily
valid at freeze out, where $x \sim 25$ and $v_X \sim \frac{1}{2}$,
but, of course, breaks down if $v_X$ is near 2.

With this parameterization, the annihilation cross section is related
to the relic density through~\cite{Kolb:1990vq}
\begin{eqnarray}
\Omega_X h^2 &\simeq& 1.07 \times 10^{9}~\gev^{-1}
\frac{n+1}{\sqrt{g_*} \mplanck} \frac{x_F^{n+1}}{\sigma_0}
\label{omegacross} \\
\sigma_0 &=& \frac{1}{c^2-1}\sqrt{\frac{8}{45}}\frac{2\pi^3}{g}
\frac{g_*^{1/2} x_F^{n+1/2}}{m_X \mplanck}e^{x_F} \, ,
\label{crossann}
\end{eqnarray}
where $x_F=m_X/T_F$, $g_*$ is the effective number of relativistic
degrees of freedom at freeze out, $g$ is the number of $X$ degrees of
freedom, and $c$ is defined by $Y(x_F) \equiv cY_{\text{EQ}}(x_F)$,
with $Y(x)$ the $X$ number density per entropy density and
$Y_{\text{EQ}}(x)$ its value if $X$ had remained in thermal
equilibrium. We set $c^2-1=n+1$, which reproduces numerical results to
within 5\%~\cite{Kolb:1990vq}.

Given the annihilation cross section for $X$ particles, their
production rate at colliders is fixed by the principle of detailed
balance, assuming time reversal symmetry~\cite{Frazer}.  Neglecting
the $i, j$ parton masses, we find
\begin{eqnarray}
\sigma(ij \to X \bar{X}; \hat{s}) &=& 
\frac{ \eta_{ij} v_X^2 (2S_X+1)^2}{4 (2S_i+1) (2S_j+1)}
\sigma(X \bar{X}\to ij; \hat{s}) \nonumber \\
&=&\frac{ \eta_{ij} (2S_X+1)^2}
{4 (2S_i+1) (2S_j+1)} \kappa_{ij} \sigmaan v_X^{2n+1} ,
\label{sigma}
\end{eqnarray}
where $\kappa_{ij}= \sigma(X \bar{X} \to ij; \hat{s})/\sigmatot$,
$\eta_{ij}$ is $\frac{1}{2}$ if $i$ and $j$ are identical and 1
otherwise, $S$ denotes spin, and all cross sections include averaging
and summing over initial and final state spins, respectively, but do
not include color factors.  Note that \eqref{sigmatot} has been used,
and so the final expression of \eqref{sigma} is not trustworthy for
$v_X$ near 2.  Equations~(\ref{omegacross})--(\ref{sigma}) determine
the minimum production cross section, given the observed relic
density, as a function of a few parameters that characterize the
properties of $X$: its mass $m_X$ and spin $S_X$, its dominant
annihilation channel $n$, and the dynamical parameters $\kappa_{ij}$.

We now turn to the case of superWIMP dark matter produced in $L$
decays.  The $L$ lifetime is extremely long (for gravitational decays,
it is of the order of hours to months), and so $L$ particles appear as
stable charged particles in colliders.  For the LHC, with $\sqrt{s} =
14~\tev$, the minimum cross section for $L$ pair production is
\begin{eqnarray}
\lefteqn{\bar{\sigma}(pp\to L^+ L^-; s)= 
\int^{\frac{4 \mL^2}{s}\frac{1}{1-\vmax^2/4}}
_{\frac{4 \mL^2}{s}} du 
\int^1_u \frac{dx}{x} }
\nonumber\\
&& \times \textstyle{\sum_{ij}} 
\left[ f^p_{q_i}(x) f^p_{\bar{q}_j} \! \left( u/x \right)
+f^p_{\bar{q}_j}(x)f^p_{q_i} \! \left(u/x \right) \right] 
\nonumber \\
&& \times \bar{\sigma}(q_i\bar{q}_j\to L^+ L^- ; us) \ ,
\end{eqnarray}
where $f^p_i$ are proton parton distribution functions, and
\begin{eqnarray}
\bar{\sigma}(q_i\bar{q}_j\to L^+ L^- ; us) 
=  \frac{1}{N_c^2} \sum_{\text{color}} 
\sigma (q_i\bar{q}_j\to L^+ L^- ; us)
\label{sigmabar}
\end{eqnarray}
is the color-averaged parton-level cross section, where $N_c = 3$, and
the right-hand side is determined by \eqref{sigma} with $X$ replaced
by $L$.  The upper limit of integration for $u$ forces $v_L < \vmax$.
We choose $\vmax^2 = 2$, so that the parametrization of
\eqref{sigmatot} may reasonably be expected to be valid in the region
of integration, and we conservatively neglect all contributions from
$v_L > \vmax$.  We also neglect subleading contributions from
(loop-induced) $gg$ fusion and three-body processes $qg\rightarrow L^+
L^- q$.  Tevatron cross sections are determined by replacing one
proton with an anti-proton and setting $\sqrt{s} = 2~\tev$.

To distinguish the metastable $L$ signal from background, we require
that {\em both} $L^+$ and $L^-$ have pseudo-rapidity $|\eta|<2.5$ and
velocity $\beta<0.7$, so that both tracks will be detected with
ionization $-dE/dx$ more than double minimum-ionizing.  The $\beta$
and $v_X$ requirements are correlated but independent, since $\beta$
is in the lab frame, and $v_X$ is in the parton center-of-mass frame.
These cuts, together with the requirement of isolated tracks in events
free of hadronic activity, should leave the signal essentially
background-free.  The event rates depend weakly on the $\eta$ cut,
dropping by about 20\% when requiring $|\eta|<0.5$.  For $\beta <0.6$,
the event rate drops by a factor of 2 to 5, depending on $m_L$.

\begin{figure}
\resizebox{3.0 in}{!}{
\includegraphics{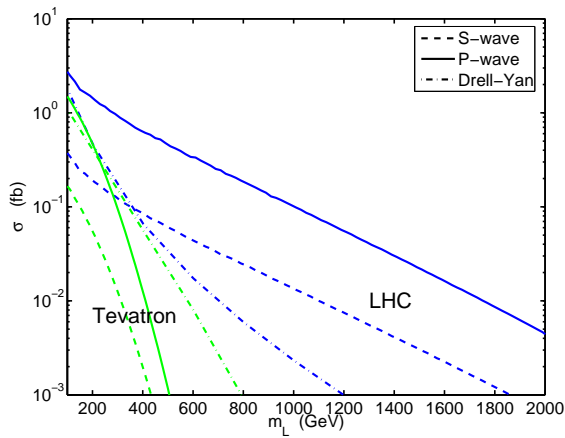}
}
\caption{Minimum Tevatron and LHC cross sections derived from
cosmology, along with the Drell-Yan cross section, for $L^+ L^-$
production in the superWIMP scenario as functions of $\mL$.  Each
event has two highly-ionizing tracks with velocity $\beta < 0.7$ and
psuedorapidity $|\eta| < 2.5$.  We assume $S_L = 0$, $\mSWIMP / \mL =
0.6$, and $\kappa_{q\bar{q}} = 0.2$ for $q=d,u,s,c,b$.
\label{fig:LHC}
\vspace*{-.2in} }
\end{figure}

In \figref{LHC} we show the minimum cross sections for $L$ pair
production at the Tevatron and the LHC as functions of $\mL$ for both
$S$- and $P$-wave annihilation, assuming scalar $L$ particles,
velocity-independent $\kappa_{q\bar{q}} = 0.2$ for $q=d,u,s,c,b$, and
$\mSWIMP / \mL = 0.6$.  As can be seen, $P$-wave annihilation in the
early Universe is suppressed by $v_L^2$ relative to $S$-wave, and so
must be compensated by larger $\sigmaan$, leading to larger minimum
collider rates.  $D$- and higher wave annihilation will imply even
larger minimum rates.  The dependence on the other parameter
assumptions is that the cross sections scale linearly with
$(2S_L+1)^2 \kappa_{q\bar{q}}$ and, to an excellent
approximation, $\mSWIMP / \mL$.  The $L$ particles are assumed to be
produced isotropically in the parton center-of-mass frame.  We have
checked that the results are insensitive to this assumption, varying
by less than 10\% for alternative distributions, such as
$\sin^2\theta$ and $(1 \pm \cos\theta)^2$.

The cosmologically constrained $L$ production cross sections shown in
\figref{LHC} are significant.  For the LHC, even for $\mL \sim
1~\tev$, cross sections as large as $0.1~\text{fb}$ are predicted for
$P$-wave annihilation.  For comparison, the Drell-Yan production cross
section $\sigma(pp \to \gamma, Z \to L^+L^-)$ is also shown in
\figref{LHC}.  At the LHC, for the parameters chosen, the minimum
cross sections we have derived typically exceed the Drell-Yan cross
sections; for $P$-wave annihilation, they are bigger by factors of 2
to 50, depending on $\mL$.  Note that these minimum cross sections are
parameter-dependent; no absolute minimum can be derived, as, for
example, these cross sections may be made arbitrarily small by taking
$\mSWIMP$ to zero.  However, if the fact that $\OmegaSWIMP h^2$ is
around $\OmegaDM h^2$ is not merely a coincidence, it is natural to
expect $\mSWIMP \sim \mL$, and it is significant that the predicted
cross sections may be large for such natural choices.

The analysis above may be adapted easily to the case of the ILC, where
the role of $\kappa_{q\bar{q}}$ is played by $\kappa_{e^+e^-}$.  For
the ILC, $\beta$ and $\vmax/2$ are nearly identical; the derivation of
collider cross sections from the relic density is therefore reliable
only when $L$ particles are produced with $\beta < 0.7$.  As examples,
consider $S_L=0$, $\mSWIMP / \mL = 0.6$, $\kappa_{e^+e^-} = 0.1$, and
$L$ particles produced with $\beta = 0.7$.  For both $\sqrt{s} =
500~\gev$ and 1 TeV, we find minimum cross sections of roughly
$4~\text{fb}$ and $60~\text{fb}$ for $S$-wave and $P$-wave
annihilation, respectively. These imply hundreds to thousands of
background-free events per year.

\begin{figure}
\resizebox{3.0 in}{!}{
\includegraphics{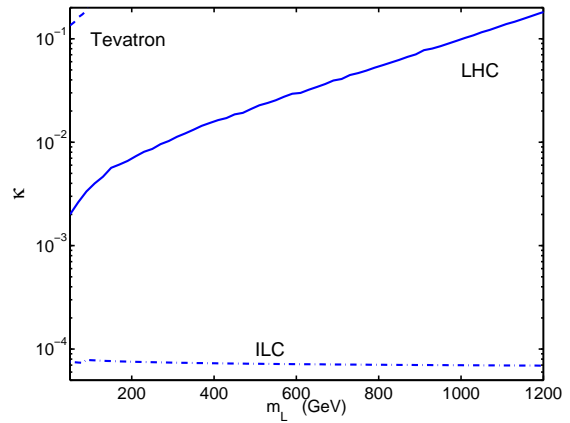}
}
\caption{Discovery limits for dark matter events in the superWIMP
scenario at the Tevatron with $L = 30~\ifb$, LHC with $L = 1~\iab$,
and ILC with $L = 1~\iab$.  We require 10 events with two doubly
minimum-ionizing tracks for discovery.  For the ILC, the beam energy
is assumed to be $\sqrt{s} \approx 2.8 \, \mL$ so that $L$ particles
are produced with $\beta = 0.7$.  We assume $S_L = 0$, $P$-wave
annihilation, and $\mSWIMP / \mL = 0.6$.  The reach in $\kappa$ scales
linearly with $(2 S_L + 1)^{-2}$ and, to an excellent approximation,
$(\mSWIMP / \mL)^{-1}$.
\label{fig:limits} 
\vspace*{-.2in} }
\end{figure}

In \figref{limits}, we give discovery limits in the $(\mL, \kappa)$
plane, where $\kappa = \kappa_{q \bar{q}}$ for the Tevatron and LHC,
and $\kappa = \kappa_{e^+e^-}$ for the ILC.  Given the cuts described
above, we expect the signal to be background-free, and so require 10
signal events for discovery.  Even assuming a highly optimistic
luminosity, the Tevatron can see the minimum dark matter signal only
for $\mL \alt 100~\gev$ and near maximal $\kappa_{q \bar{q}}$.  The
LHC does much better, probing $\kappa_{q \bar{q}} > 3 \times 10^{-3}$
for $\mL \sim 100~\gev$, and $\mL < 1.2~\tev$ for $\kappa_{q \bar{q}}
\sim 0.2$.  Finally, the ILC will provide phenomenal coverage down to
$\kappa_{e^+e^-} \sim 10^{-4}$ for all $\mL$. The process $L^+ L^- \to
e^+e^-$ may be absent at tree-level; in fact, all processes $L^+ L^-
\to f \bar{f}$ may be suppressed if annihilation to Higgs bosons
dominates.  However, even in these cases, since $L$ must be weakly
coupled to {\em some} standard model particles to explain $\Omega_L
h^2 \sim \OmegaDM h^2$, we expect $\kappa_{e^+e^-} \sim 10^{-3}$ to be
generated at loop-level.  Of course, this requires kinematically
accessible $L$ pairs.  This is reasonable for the lower range of $\mL$
plotted, but requires later stages of the ILC program for the upper
range.

We now consider WIMP dark matter scenarios.  WIMP pair production is
invisible at colliders.  Here we consider the monojet signal of WIMP
pairs produced in association with a gluon or quark.  The cross
section for $pp \to X \bar{X} j$ is not simply related to the cross
section for $pp \to X \bar{X}$.  However for collinear or soft jets,
these rates are related by splitting functions~\cite{Birkedal:2004xn}.
The color-averaged parton-level differential cross sections are, then,
\begin{eqnarray}
\lefteqn{\frac{d}{dz \, d\!\cos\theta}
\bar{\sigma}(q(\bar{q}\to \bar{q}g) \to X\bar{X}g; \hat{s})} 
\nonumber \\
&\approx& F_{\bar{q} \to g}(z, \theta) \, 
\bar{\sigma}(q\bar{q}\to X\bar{X};(1-z)\hat{s}) 
\label{fact1} \\
\lefteqn{\frac{d}{dz \, d\!\cos\theta}
\bar{\sigma}(q(g\to \bar{q}q) \to X\bar{X}q;\hat{s})} 
\nonumber \\
&\approx& F_{g\to q}(z, \theta)  \frac{2S_q+1}{2S_g+1}
\bar{\sigma}(q\bar{q}\to X\bar{X}; (1-z)\hat{s}) \ ,
\label{fact2}
\end{eqnarray}
where the splitting functions are~\cite{Halzen:1984mc}
\begin{eqnarray}
F_{q\to g}(z, \theta) &=& \frac{4}{N_c}\frac{\alpha_s}{\pi}
\frac{1+(1-z)^2}{z}\frac{1}{\sin^2\theta}\\
F_{g\to q}(z, \theta) &=& \frac{4}{N_c^2-1}\frac{\alpha_s}{\pi}
\frac{z^2+(1-z)^2}{\sin^2\theta} \ ,
\end{eqnarray}
with identical expressions for $q \to \bar{q}$.  Here $i \to jk$ means
that initial state parton $i$ radiates parton $k$, which becomes the
final state jet.  The parameter $z = E_k/E_i$ varies from 0 to $1 -
4m_{X}^2 / [s (1 - v_{\text{max}}^2/4)]$ and $\theta$ is the angle
between $i$ and $k$; both are defined in the parton center-of-mass
frame.  Given these parton-level results, the LHC color-averaged
differential cross section is
\begin{eqnarray}
\lefteqn{\frac{d}{dz \, d\!\cos\theta}
\bar{\sigma}(pp\to X\bar{X}g, X\bar{X}q, 
X\bar{X}\bar{q};s)}
\nonumber\\
&&\approx 
\int^{\frac{4m_{X}^2}{(1-z)s(1-v_{\text{max}}^2/4)}}
_{\frac{4m_{X}^2}{(1-z)s}} du 
\int^1_u \frac{dx}{x} 
\sum_{i,j=q,\bar{q},g}^{i \ne j} f^p_i(x) f^p_j(u/x) \nonumber\\
&&
\times \frac{d}{dz d\cos\theta}\bar{\sigma}(i(j\to
 \bar{i}k) \to X\bar{X}k; us) \ .
\end{eqnarray}

\begin{table}[tb]
\caption{The minimum monojet signal $S$ from $pp \to X \bar{X} j$
  at the LHC in the WIMP dark matter scenario.  We assume scalar WIMPs
  with mass 100 GeV and $P$-wave annihilation and require jets to have
  $\sin\theta_{\text{lab}} > 0.1$ and $p_T^{\text{min}}$ as indicated.
  Also given are the standard model background $B$ and the
  significance $S/\sqrt{B}$, assuming integrated luminosity $1~\iab$.
\label{table:I} }
\begin{tabular}{cccc}
$p_T^{\text{min}}$ \qquad & \qquad $S$ \qquad  
& \qquad $B$ \qquad & \qquad $S/\sqrt{B}$ \qquad \\ \hline
30 GeV \rule[0mm]{0mm}{4mm} \qquad & \qquad 18.6 fb \qquad 
& \qquad 1300 pb \qquad & \qquad 0.51 \qquad \\
100 GeV \rule[0mm]{0mm}{0mm} \qquad & \qquad 4.1 fb \qquad 
& \qquad 130 pb \qquad & \qquad 0.36 \qquad \\ \hline
\end{tabular}
\end{table}

To ensure that the monojet events are detectable, we require the jets
to have $\sin\theta_{\text{lab}}>0.1$ and $p_{T} > p_T^{\text{min}}$.
The factorization of \eqsref{fact1}{fact2} holds formally only in the
limit of soft or collinear jets, but it has been shown to be
reasonably accurate even away from these limits in the region we have
included~\cite{Birkedal:2004xn}.  In contrast to the superWIMP case,
where the background is negligible, monojet events in the WIMP
scenario suffer from a huge irreducible background from $pp \to \nu
\bar{\nu} j$.  At the ILC, the analogous single photon signal may be
improved by an additional cut on the photon
energy~\cite{Birkedal:2004xn}.  Unfortunately, this approach is not
effective at the LHC because the parton center-of-mass energy is not
fixed.  Table~\ref{table:I} gives cross sections for the monojet
signal for $\mWIMP = 100~\gev$ and two values of $p_T^{\text{min}}$.
The background, with the identical cuts implemented using the
simulation package COMPHEP~\cite{COMPHEP}, is also given.  Although
thousands of WIMP monojet events are expected given integrated
luminosity $1~\iab$, the overwhelming background leads to very small
$S/\sqrt{B}$, making discovery extremely difficult.

In summary, if stable WIMPs or superWIMPs exist, requiring that they
not overclose the Universe implies efficient dark matter production
rates at colliders.  Using model-independent parametrizations, we have
determined lower bounds on dark matter production rates as functions
of a few parameters characterizing WIMP and superWIMP properties.  For
WIMP dark matter, the $X \bar{X} j$ signal is swamped by background.
On the other hand, for natural parameters in the superWIMP scenario,
the derived rate for the production of two metastable charged
particles may be much larger than the Drell-Yan cross section and
implies spectacular signals at the LHC and, if kinematically
accessible, the ILC.  These superWIMP results imply promising
prospects not only for detection of dark matter signals, but also for
detailed studies~\cite{Hamaguchi:2004df} of dark matter at future
colliders.

{\em Acknowledgments} --- We thank M.~Perelstein for useful comments.
The work of JLF is supported in part by NSF CAREER grant
No.~PHY--0239817, NASA Grant No.~NNG05GG44G, and the Alfred P.~Sloan
Foundation.

\vspace*{-.2in}



\end{document}